\documentclass[superscriptaddress,groupedaddress]{revtex4}  
\usepackage{graphicx}  
\usepackage{dcolumn}   
\usepackage{bm}        
\usepackage{amssymb}   
\usepackage{amsmath}
\usepackage{feynmp}    
\begin{document}
\newcommand{\bq}{\begin{equation}}
\newcommand{\eq}{\end{equation}}
\newcommand{\bqn}{\begin{eqnarray}}
\newcommand{\eqn}{\end{eqnarray}}
\newcommand{\nb}{\nonumber}
\newcommand{\lb}{\label}

\widetext

\title{Primordial Neutrinos:\\Hot in SM-GR-$\Lambda$-CDM,\\Cold in SM-LGT}
%
\author{Ahmad Borzou} \affiliation{EUCOS-CASPER, Physics Department, Baylor University, Waco, TX 76798-7316, USA}\email{ahmad_borzou@baylor.edu}
%
\vskip 0.25cm
       
\date{\today}

\begin{abstract}
We replace general relativity (GR) and the cosmological constant ($\Lambda$) in the standard cosmology (SM-GR-$\Lambda$-CDM) with a Lorentz gauge theory of gravity (LGT) and show that the standard model (SM) neutrinos can be the cold dark matter (CDM) because (i) the expansion of the universe at early times is not as sensitive to the amount of radiation as in the SM-GR-$\Lambda$-CDM and (ii) in LGT there exists a spin-spin long-range force that is very stronger than the Newtonian gravity and interacts with any fermion including neutrinos. Assuming that neutrinos as heavy as 1eV are the cold dark matter, the lower bound on the dimensionless coupling constant of LGT is derived to be $10^{-7}$ which is small enough to be consistent with the upper bound that can be placed by the electroweak precision tests. We also show that the vacuum energy does not gravitate in LGT and a decelerating universe shifts spontaneously to an accelerating one right at the moment that we expect. Therefore, current observations can be explained in our cosmological model (SM-LGT) with lesser assumptions than in the SM-GR-$\Lambda$-CDM.
\end{abstract}

\maketitle

\section{Introduction}
In 1937 Zwicky found the evidence for existence of dark matter through measurement of the masses of large galaxy clusters. The existence of such matter was later re-confirmed through observation of rotation curves of galaxies and more reliably through the CMB anisotropies. Despite 80 years of appreciation of dark matter, it is still not known what particles are constituting it. The experimental searches for them have not been successful \cite{XENON2017,LUX2017,PandaX2017,Kahlhoefer2017,CMS2013,ATLAS2012}. In 1980s neutrinos were thought to be a plausible dark matter candidate. The idea was later abandoned because a hot dark matter can not explain the galaxy formations and neutrinos were hot according to the standard cosmology. Nevertheless, hotness of neutrinos is only a prediction of the SM-GR-$\Lambda$-CDM model which has not been confirmed through independent experiments. 

Some cosmological observations are only an indication of a certain expansion profile. For example an accelerating expansion in the late times is needed in order to explain the observations regarding the Supernova type Ia. Or, a certain type of expansion is needed at early times to explain the observations of the primordial light elements. These observations therefore can be used to draw a theory-independent experimental expansion profile for the universe. 
Almost any cosmological model can acquire this expansion profile by tweaking the matter, the radiation and the vacuum energy in the universe. A successful model therefore is not the one that has the required expansion profile but instead is the one whose assumptions---predictions---regarding the matter, the radiation, and the vacuum energy are confirmed through other independent observations. In the standard cosmology this expansion profile can be reached only if neutrinos are hot and the vacuum energy has a positive but nearly zero value. Neither of which are confirmed by independent observations. In fact the latter is so fine-tuned that is seen as one of the greatest challenges of the theoretical physics \cite{Weinberg89}.

In this paper we replace GR and $\Lambda$ of the standard cosmology by a single-parameter Yang-Mills theory of gravity based on the Lorentz symmetry of spinors---LGT---and show that the cold dark particles can be the neutrinos in the standard model of particle physics. We show that first, the early-times expansion of the universe is not very sensitive to the amount of radiation, i.e. the model gets the required early-times expansion without tweaking the amount of radiation in the universe. Second, at late times, a decelerating universe spontaneously shifts to an accelerating one, i.e. it is not necessary to fine-tune the magnitude of the cosmological constant in order to reach the needed expansion profile. Therefore, this model only needs an input about the cold matter---whose value is obtained through independent observations.

In comparison with the standard cosmology, the model needs one less assumption regarding the vacuum energy and one less assumption regarding the radiation, in order to reach the same expansion profile. Moreover, while in SM-GR-$\Lambda$-CDM a new theory for the dark particles is needed, in SM-LGT neutrinos are a serious dark matter candidate because (i) there is no need to assume that they are hot and (ii) LGT provides a long-range spin-spin force---that interacts with any spin-half-integer particle---that can keep neutrinos in thermal contact until they are non-relativistic. In this paper we show that if neutrinos have a mass of 1eV and the dimensionless coupling constant of LGT is 6 orders of magnitude smaller than the electroweak couplings, cold neutrino dark matter is a viable scenario and the spin-spin force has negligible effects in the electroweak precision tests.

\section{Lorentz gauge theory of gravity}
The kinematics of LGT is the same as the version of general relativity in which fermions are also accounted for and is precisely described in \cite{WeinbergGRBook}. In this kinematics the geometry is Riemannian and no torsion is added to the space-time. Fermions are introduced on the tangent spaces defined by a set of four orthonormal unit vectors called the tetrad. The theory is invariant not only under any general coordinate transformation but also under any local Lorentz transformation of spinors \cite{WeinbergGRBook}. Each of the two symmetries introduce a conserved current. The conserved current associated with the space-time symmetry is the energy-momentum tensor while the one associated with the Lorentz invariance of spinors is called the Lorentz current and is derived in \cite{Borzou4}. Any of the two currents can be used to build a theory of gravity. In general relativity the energy-momentum is the source of gravity while in LGT the Lorentz current generates the gravitational fields. The source in a given theory determines the dynamical variable of it. When the energy-momentum is the source of gravity, the metric is the dynamical variable. When the Lorentz current is the source of gravity, the spin connections are the dynamical variables. Therefore, with the same kinematics and starting with the same Lagrangian, we can derive two different theories---field equations---depending on the chosen source.

Lorentz current and not the energy-momentum tensor being the source of gravity has several advantages. First, the coupling constant of the theory becomes dimensionless  which is the first renormalizability condition. Second, since the vacuum energy couples to gravity through the determinant of the metric and since the metric is not dynamic in LGT, vacuum energy does not go into the source of LGT and therefore does not gravitate \cite{Borzou4}. This solves the conflict with quantum field theory. Later in this paper we show how LGT explains the accelerating expansion of the universe without a dark energy. Third, when the metric is not dynamic, time is also not discrete and therefore there is no conflict with quantum mechanics anymore. It should also be mentioned that predictions of LGT are not easily manipulable since it only has one free parameter. 

The Lagrangian and the filed equations of LGT can be found in \cite{Borzou1}. For the sake of comparison with general relativity, we are more interested in macroscopically effective field equations which are reached after integrating a Planck length out of the full theory \cite{Borzou2} 
\bqn
\lb{eq:fieldequation}
\nabla^{\nu}R_{\mu \nu \rho \sigma} = 4\pi G \left( \nabla_{\sigma}T_{\mu\rho}-\nabla_{\rho}T_{\mu\sigma} \right),
\eqn
where $\nabla$ refers to the covariant derivative in the Riemannian geometry, $R$ is the curvature tensor, and $T$ is the fermionic part of the energy-momentum tensor---bosons do not contribute to the Lorentz current and hence do not gravitate in LGT \cite{Borzou4}. Also, it should be emphasized that LGT is not a higher derivative theory as this equation contains only 2 derivatives on the dynamical variable of the theory. A contraction of the equation reads 
\bqn
\nabla^{\nu}R_{\nu \sigma} = 4\pi G  \nabla_{\sigma}T,
\eqn
which can be compared with the similar equation in general relativity
\bqn
\nabla^{\nu}R_{\nu \sigma} = -4\pi G  \nabla_{\sigma}T.
\eqn
Therefore, for highly symmetric space-times in which only a few parameters are not determined by the symmetries---Schwarzschild and de Sitter for example---LGT and general relativity both share the same exact solutions in vacuum \cite{Borzou1}. Also, inside matter in the linear regions the two theories are exactly Newtonian \cite{Borzou2}. The minus sign difference is needed for a correct boundary condition and is because the dynamic variable in LGT is proportional to the derivative of the daynamic variable in GR.

The Newtons gravitational constant in LGT is given in terms of the coupling constant of LGT $g$ via $G=g\delta ^2/\left(40\pi\right)$ after the Planckian length $\delta$ is integrated out of the full theory \cite{Borzou2}. An analogous case is when the W boson's mass is integrated out of the electroweak theory and the Fermi constant is emerged in terms of the coupling constant of the underlying theory $G_{\text{F}}=\sqrt{2}g_{\text{w}}^2/\left(8 m^2_{\text{W}}\right)$. In both cases the dimensionless coupling constant of the underlying theory is divided by the square of a mass. This is the reason why Fermi constant is extremely smaller than the coupling constant of the electroweak theory. In the same way we can conclude that the coupling constant of LGT is way stronger than the Newtonian constant something that is crucial to our cold neutrino dark matter proposal that follows next. 

Before moving forward it should be mentioned that whenever that it is possible, LGT uses the ideas and the rich literature of the Poincare gauge theories (PGT). Nevertheless, there are differences that motivate this research. Perhaps the most basic difference is in the source of gravity in each theory. While in PGT energy-momentum tensor is one of the sources of gravity, in LGT the Lorentz current is the source. Consequently, PGTs are not renormalizable---their coupling constant still has -2 dimension---while LGT is power-counting renormalizable with a dimensionless coupling constant. Another difference is that in the 2-derivative versions of PGTs, the spin-generated field does not propagate. Higher derivative terms are needed to solve the issue. But, higher derivative theories usually propagate Ostrogradsky ghosts and violate unitarity. See the description in for example \cite{Horava2009}. Moreover, higher derivative PGTs usually have a potential that linearly increases with the distance form the source \cite{SIJACKI1982435,ASNA:ASNA2103020310}. Also, it remains to be answered why most of theories in physics have only 2 derivatives on their dynamical variable but a gravitational theory should have 4 derivatives. LGT on the other hand is a 2-derivative theory where both the mass- and the spin-generated fields propagate. It also does not contain any potential that linearly increases with the distance from the source \cite{Borzou2}.

In LGT mass and spin of fermions both gravitate. The mass-generated force is the well-known Newtonian gravity whose strength is determined by Newtons gravitational constant---the coupling constant of LGT multiplied by the square of a very small length---as discussed above. The spin-generated force has not been observed and its strength is determined with the coupling constant of LGT alone. As was discussed above, this latter force is very stronger than the Newtonain gravity. It remains to mention that the force is absent in classical systems since they do not have a net intrinsic spin. Moreover, since any half-integer-spin particle interacts with this force, and in order to pass the precision tests of the electroweak theory, the coupling constant of LGT should be smaller than those in the standard model. In the following we show that it can be 6 orders of magnitude smaller than the coupling constants of the electroweak and yet keep neutrinos in thermal contact until they are cold. The Feynman rules for the theory after neglecting the curvature of the space-time are derived in \cite{Borzou4,Borzou1}. The graviton propagator is given by      
\bqn
\begin{fmffile}{propagator}~~~~~~~~~~
  \parbox{5mm}{\begin{fmfgraph*}(40,2)
    \fmfleft{l}
    \fmf{wiggly}{l,r}
    \fmfright{r}
    \fmflabel{$ij\nu$}{r}
     \fmflabel{$mn\mu$}{l}
     \end{fmfgraph*}}
\end{fmffile}
~~~~~~~~~~~~~= \frac{1}{2}\eta_{\mu\nu}\left(\eta_{mi}\eta_{nj}-\eta_{mj}\eta_{in}\right)/p^2,
\eqn
There are only two self-interaction vertices in LGT. However, in the tree level only one of them is important \\\\\\
$~~~~~~~~~~
    \begin{fmffile}{SelfInteraction}
      \setlength{\unitlength}{.5cm}
      \parbox{30mm}{\begin{fmfgraph*}(3.5,3.5)
  \fmfright{i1,i2}
  \fmfleft{o}
  \fmf{boson,label.side=right}{i1,v}
  \fmf{boson}{o,v}
  \fmf{boson}{i2,v}
  \fmflabel{$i_2j_2\mu_2$}{i1}
  \fmflabel{$i_3j_3\mu_3$}{i2}
  \fmflabel{$i_1j_1\mu_1$}{o}
  \end{fmfgraph*}}
  \end{fmffile}
$\\\\\\which is equivalent to\\
$-g\eta^{i_1i_3}\eta^{j_1j_2}\eta^{i_2j_3}\Big(2k_3^{\mu_2}\eta^{\mu_3\mu_1}-k_3^{\mu_1}\eta^{\mu_2\mu_3}-2k_2^{\mu_3}\eta^{\mu_2\mu_1}+k_2^{\mu_1}\eta^{\mu_2\mu_3}\Big)/4 + \text{permutations of corresponding i and j}$.
Finally assuming that the matter is described by a Dirac Lagrangian, the only relevant diagram at the tree level is a purely axial vector describing the interaction between a fermion and a spin-generated graviton
\begin{fmffile}{spin}
 \bqn
 \lb{eq:VertexDiag}
 \setlength{\unitlength}{.5cm}
 \parbox{20mm}{\begin{fmfgraph*}(3.5,3.5)
  \fmfleft{i1,i2}
  \fmf{fermion}{i1,v,i2}
  \fmf{boson}{o,v}
   \fmfright{o}
   \fmflabel{$mn\mu$}{o}
 \end{fmfgraph*}}
~~~~~~=\frac{ig}{4}\gamma^k\gamma^5\varepsilon_{kmn\mu},
 \eqn
\end{fmffile}
where epsilon is the Levi-Civita symbol.

\section{Zero order cosmology in LGT}
A Friedmann-Lemaitre-Robertson-Walker space-time in LGT is studied in \cite{Borzou3}. After substituting the metric into Eq.~(\ref{eq:fieldequation}) the following differential equation is reached for a flat universe
\bqn
\dddot{a}-2\frac{\dot{a}^3}{a^2}+\frac{\dot{a}\ddot{a}}{a}=4\pi G\frac{(1-3w)(1+w)\rho_{0}}{a^{3(1+w)}}\dot{a} ,
\eqn
where dot indicates derivative with respect to time, and $w$ is the ratio of pressure to energy density. From the equation it can be found that the source term is zero for relativistic matter with $w=\frac{1}{3}$ and any matter with $w=-1$. On the other hand when the right hand side is zero---like the two mentioned cases---two exact solutions exist. One is the de Sitter space-time and the other equals the solution of GR for radiation dominated universe. Therefore, the solutions of general relativity in the presence of dark energy or radiation are the vacuum solutions of LGT. An interesting observation is that in this equation the derivative of the acceleration of a decelerating universe in the absence of matter---the conditions for the universe at the late times--- is positive. That means the acceleration will eventually change sign with no need for a dark energy. In order to see when this transition happens, the differential equation is numerically solved assuming that the cold matter density in the universe is $0.3$ of the critical density and $\dot{a}$ and $\ddot{a}$ at $a=1$ have the same values as in the standard cosmology. We also have made no assumption regarding the values of vacuum energy and radiation as they will have no contribution to the equation. The results of the numerical study is presented in \cite{Borzou3}. It shows that in LGT the universe, spontaneously shifts its expansion mode at two points. And, the two points coincide with when the standard cosmology shifts from radiation-dominated to matter-dominated era and when the decelerating universe jumps into an accelerating one. At the former point, the standard cosmology puts a stringent constraint on the neutrino radiation. While at the latter point, it requires a fine-tuned cosmological constant. None of which are confirmed through independent observations. Since LGT puts no severe constraint on the neutrino radiation, one of the conditions regarding a cold neutrino dark matter scenario is satisfied.

\section{Neutrino-Graviton scattering}
For the cold neutrino dark matter scenario, the most important interaction that is allowed in LGT is the scattering of spin-generated gravitons off the primordial neutrinos at low energies. The amplitude can be inferred by the following diagrams
\begin{widetext}
\begin{eqnarray}
~\nonumber\\
    &&\begin{fmffile}{Nu_Graviton}
      \setlength{\unitlength}{1.cm}
      \parbox{50mm}{\begin{fmfgraph*}(5,3)
        \fmfleft{ld,lu}
        \fmfright{rd,ru}
        \fmf{fermion,label=$\nu\left(p\right)$}{ld,w1}
        \fmf{fermion,label=$\nu\left(p'\right)$}{w0,lu}
        \fmf{fermion,label=$p+k$}{w1,w0}
        \fmf{boson,label=$A\left(k\right)$}{w1,rd}
        \fmf{boson,label=$A\left(k'\right)$}{ru,w0}
      \end{fmfgraph*}
      }
      +      
      \parbox{50mm}{\begin{fmfgraph*}(5,3)
        \fmfleft{ld,lu}
        \fmfright{rd,ru}
        \fmf{fermion,label=$\nu\left(p\right)$}{ld,wl}
        \fmf{boson,label=$A\left(k\right)$,label.side=right}{wr,rd}
        \fmf{phantom}{wl,lu}
        \fmf{phantom}{wr,ru}
        \fmf{fermion,label=$p-k'$}{wl,wr}
        \fmf{fermion,tension=0,label=$\nu\left(p'\right)$,label.side=left}{wr,lu}
        \fmf{boson,tension=0,label=$A\left(k'\right)$,label.side=right}{wl,ru}
      \end{fmfgraph*}
      }
      +
      \parbox{50mm}{\begin{fmfgraph*}(5,3)
        \fmfleft{ld,lu}
        \fmfright{rd,ru}
        \fmf{fermion,label=$\nu\left(p\right)$}{ld,wl}
        \fmf{boson,label=$A\left(k\right)$,label.side=right}{wr,rd}
        \fmf{boson,label=$p-p'$}{wl,wr}
        \fmf{fermion,label=$\nu\left(p'\right)$,label.side=right}{wl,lu}
        \fmf{boson,label=$A\left(k'\right)$,label.side=left}{wr,ru}
      \end{fmfgraph*}
      }
    \end{fmffile}.\nonumber\\
\end{eqnarray}
\end{widetext}
The exact form of the amplitude depends very much on the neutrino model which is still an active area of research. But, since amplitudes are dimensionless, the results from different neutrino models should stand within the same order of magnitude and should not change the overall conclusion that will be drawn in this paper. 
If neutrinos are simple Dirac particles, at very low energies the amplitude square reads
\bqn
\sum_{\text{spin}}|{\cal{M}}|^2  
&=&\left(\frac{g}{4}\right)^4\Bigg(44 \sin (\theta )-60 \sin (2 \theta )\nb\\
&+&28 \sin (3 \theta )-6 \sin (4 \theta )-76 \cos (\theta )\nb\\
&+&60 \cos (2 \theta )-4 \cos (3\theta )-8 \cos (4 \theta )\nb\\
&+&20 + {\cal{O}}\left(\frac{1}{m_{\nu}}\right)\Bigg),
\eqn 
where $m_{\nu}$ is the neutrino mass and $\theta$ is the angle between graviton's final and initial directions of motion. The total cross-section in terms of the calculated amplitude is given by 
\bqn
\lb{eq:totalXS}
\sigma_{\text{T}}
=\int_{0}^{\pi}d\theta\sin\left(\theta\right)\frac{\left(1+{\cal{O}}\left(\frac{1}{m_{\nu}}\right)\right)}{2^7\pi m_{\nu}^2}
\sum_{\text{spin}}|{\cal{M}}|^2
\simeq 10^{-3}\frac{g^4}{m_{\nu}^2}.
\eqn
In the following this will be used in order to find the neutrino freeze-out temperature.

\section{Cold neutrino dark matter scenario}
The upper bound on the neutrino mass comes from a combination of different experiments and is $\sim$1eV \cite{ParticleData2017}. That means the neutrino freeze-out has to be near or below this value in order for them to be cold. If neutrinos were only interacting through the weak and the classical gravitational fields, they would freeze-out at temperatures around 1MeV. In LGT however there is a long-range spin-spin force that interacts with any fermion so that the Lorentz symmetry is locally preserved and as was discussed above is very stronger than the classical gravitational force. The upper bound on the force comes from the electroweak precision tests. Therefore, the force can not be as strong as those in the standard model. We now would like to find a lower bound on its strength assuming that it keeps neutrinos in thermal contact until they are cold. The cold neutrino dark matter will be a viable scenario only if the lower and the upper bounds that are coming from independent considerations leave any possible phase space for the coupling constant of the force.

Keeping the upper bound on the strength of the spin-spin force in mind, its effects at temperatures above 1MeV can be neglected. When temperature falls below this value, scattering of spin-generated gravitons off the neutrinos will be the dominant interaction and as far as its collision time is less than the cosmic time, neutrinos remain in thermal contact. The collision time can be calculated through 
\bqn
t_{\text{c}}\simeq 1/\left(\sigma_{\text{T}}n_{\nu}v_{\text{rel}}\right), 
\eqn
where $v_{\text{rel}}$ is the relative speed between neutrinos and gravitons and is equal to one, $\sigma_{\text{T}}$ is the total cross-section given by Eq.~(\ref{eq:totalXS}) and $n_{\nu}$ is the neutrino---dark matter in our scenario---number density at the time of freeze-out and can be derived in terms of the CMB temperature using the dark matter density at the present time 
\bqn
n_{\nu}=\rho_{_{0\text{DM}}}\text{T}^3/\left(m_{\nu}\text{T}_0^3\right),
\eqn
where naught refers to the present time. The collision time therefore reads 
\bqn
t_{\text{c}}\simeq 10^3m_{\nu}^3\text{T}_0^3/\left(\rho_{_{0\text{DM}}}g^4\text{T}^3\right). 
\eqn
\begin{figure}
\includegraphics[width=70mm,scale=1.]{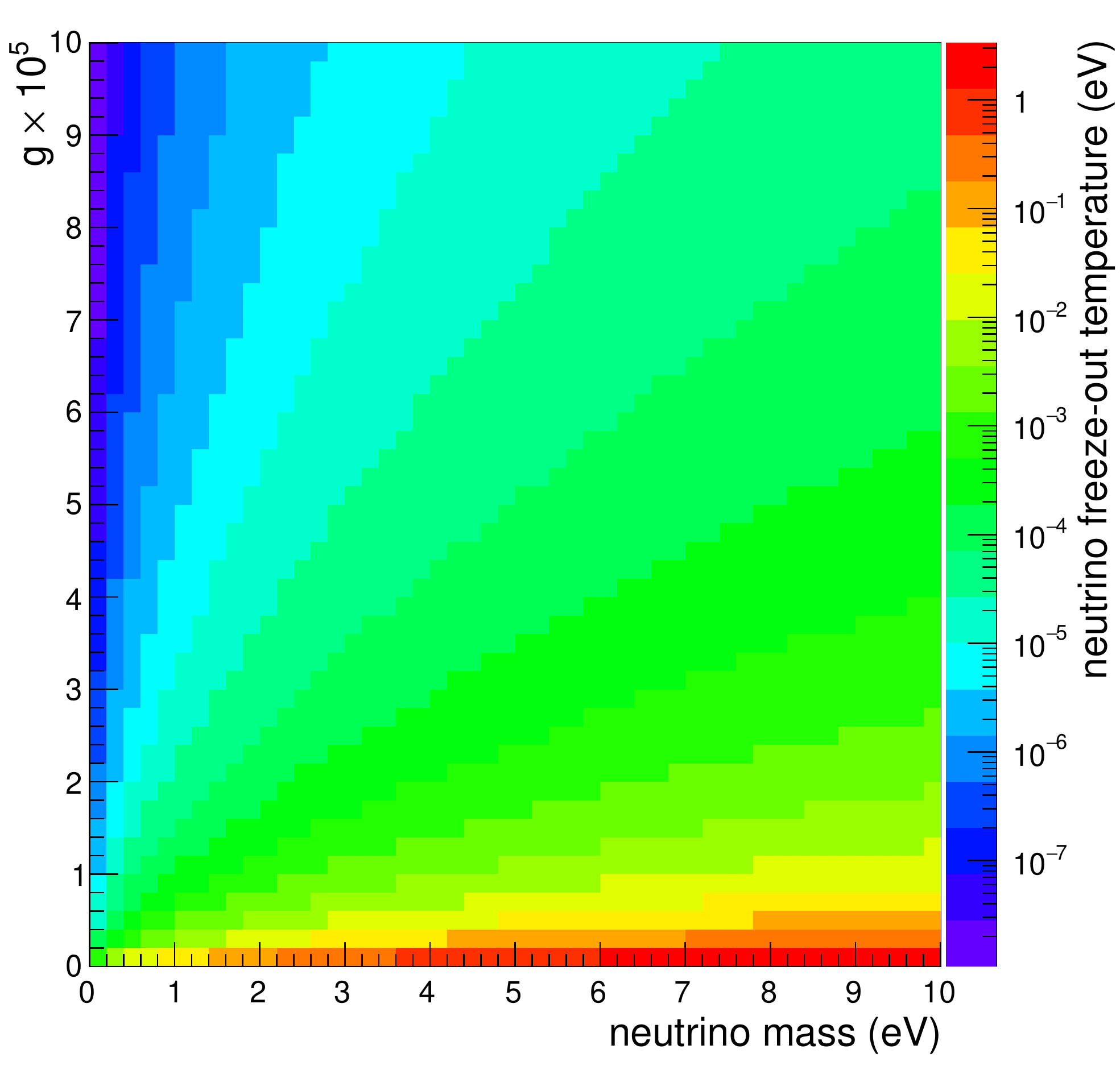}
\caption{\label{fig:freezeout} The neutrino freeze-out temperature as a function of neutrino mass and the coupling constant of LGT. Neutrinos will be non-relativistic at the freeze-out in the drawn phase space.}
\end{figure}
The cosmic time is the inverse of the Hubble parameter which itself is given by the observation-based expansion profile of the universe, i.e. it has to be the same in every model of cosmology at the cost of tweaking matter, radiation and vacuum energy. In \cite{Borzou3} it is shown that within the accuracy of the analysis, LGT also reproduces the same expansion profile. For temperatures below a few electron-volts the cosmic time is given by 
\bqn
t_{\text{H}}=\sqrt{3/\left(8\pi G m_{\nu}n_{\nu}\right)},
\eqn
assuming that neutrinos are the dark matter that constitute the whole matter in the universe. The freeze-out temperature will be obtained by equating the latter two times
\bqn
\text{T}_{\text{freeze}} = \frac{0.1m_{\nu}^2}{\left(10^6 g\right)^{\frac{8}{3}}} \left(\text{eV}^{-1}\right),
\eqn
which is drawn in Fig.~\ref{fig:freezeout}.  
The lower bound on the coupling constant of the spin-spin force can be reached by inserting $\text{T}_{\text{freeze}} \lesssim m_{\nu}$ into the latter equation
\bqn
g \gtrsim 10^{-6}\left(0.1m_{\nu}\text{eV}^{-1}\right)^{\frac{3}{8}}.
\eqn
If neutrinos are as heavy as 1eV, the smallest value for the coupling constant is $g \sim 10^{-7}$. This coupling constant is nearly six orders of magnitude smaller than those of the electroweak---it is small enough to be neglected in the precision tests of the electroweak---but still extremely larger than that of the Newtonian force of gravity.

\section{Neutrino Bound States}
The similarity between the vertex diagram in Eq.~(\ref{eq:VertexDiag}) and the only interaction in QED indicates that a gas of neutrinos behave in a very similar way as a gas made of electrons, positrons, and protons. We are specially interested in the non-relativistic behavior of such system and therefore need to derive the spin-spin force of LGT in non-relativistic limits. The corresponding potential energy in the momentum space can be found through the following diagram for two distinguishable fermions
\begin{eqnarray}
~\nonumber\\
    &&\begin{fmffile}{Spin_Spin_Force}
      \setlength{\unitlength}{1.cm}
      \parbox{50mm}{\begin{fmfgraph*}(5,3)
        \fmfleft{ld,lu}
        \fmfright{rd,ru}
        \fmf{fermion,label=$p$}{ld,wl}
        \fmf{fermion,label=$k$,label.side=left}{rd,wr}
        \fmf{boson,label=$p-p'$}{wl,wr}
        \fmf{fermion,label=$p'$,label.side=right}{wl,lu}
        \fmf{fermion,label=$k'$,label.side=left}{wr,ru}
      \end{fmfgraph*}
      }
    \end{fmffile},\nonumber
\end{eqnarray}
which is equivalent to
\bqn
{\cal{M}} = \frac{-g^2}{16}\bar{u}^{s}\left(p'\right)\gamma^k\gamma^5 u^{s}\left(p\right)\bar{u}^{r}\left(k'\right)\gamma_k\gamma^5 u^{r}\left(k\right)/\left(p-p'\right)^2.
\eqn
The leading term reads ${\cal{M}} = \pm \frac{4m^2g^2}{16|\vec{p}-\vec{p}'|^2}$ where the plus and minus signs are for when the spin of the two interacting particles are in opposite and same directions correspondingly. The potential energy in the configuration space is the Fourier transform of this latter amplitude and reads $V\left(r\right)=\mp \left(\frac{g}{4}\right)^2/\left(4\pi r\right)$. As expected, the spin-spin force is the same as the electric force if the electric charge $e$ is replaced by $\frac{g}{4}$ and is attractive for spins in opposite directions and repulsive for same direction spins. It is now quite easy to predict the possible phenomena in LGT by looking at whatever that is possible in QED and make the $e \rightarrow \frac{g}{4}$ replacement. Bound states---atoms---made of two neutrinos with opposite spins is the first to expect. The energy spectra is the same as in positronium with mass of electron replaced by mass of neutrino and the electric charge replaced by the coupling constant of LGT and reads $E_n = - m_{\nu}\left(\frac{g}{4}\right)^4/\left(4n^2\right)$. If energies are low enough that neutrinos can not escape the ground state of the neutrino-neutrino system, the bound state will be formed just like what takes for a hydrogen atom to form.
Since spins are opposite in neutrino-neutrino atoms, the spin of the whole system is zero. This is crucial to our cold neutrino dark matter proposal as we will see below. 

The question to be asked at this point is when these bound states form. When neutrinos get cold, their temperature is near 1eV and this is much higher than the energy of the ground state of the neutrino-neutrino system. On the other hand, the fast expansion of the universe freezes out the neutrinos before their temperature is low enough to form the bound states. Nevertheless, since neutrinos are cold at the freeze-out, their masses contribute into the generation of the mass-generated---Newtonian---gravity in the same way that we expect from any cold dark matter candidate, i.e. our dark matter has formed at temperatures above 1eV.  

According to the structure formation theory, as time passes the perturbations in the matter content of the universe---mostly cold dark matter in the form of neutrinos in our model---grow under attractive gravitational forces. At around the time of galaxy formation when enough visible and invisible matter is trapped under their own gravitational attractions, the system is not expanding anymore. It is a gas of mostly neutrinos as the invisible matter and other visible particles. This is when the neutrino-neutrino bound states start to form. The reason is that the spin-spin force is tens of orders of magnitude stronger than the Newtonian force and is long-range and the system is not expanding to effectively switch off the intractions. At this point, due to the interactions that are governed by the vertex diagram in Eq.~(\ref{eq:VertexDiag}), the system starts to cool off by radiating away soft spin-generated gravitons. This is just similiar to when a gas of electrons and protons are trapped in a non-expanding box and cool off through the balck body radiation.
The reason our confined neutrino gas can radiate energy away is that the vertex in Eq.~(\ref{eq:VertexDiag}) allows to draw a Feynman diagram like the Bremsstrahlung radiation in QED. In other words, a decelerating neutrino radiates a soft spin-generated graviton and cool down. When neutrinos in our pre-galaxy system are cooled enough, they will trap in the bound states and make spin 0 systems. Note that a composite spin 0 system like the hydrogen atom is boson as far as the energies of the particles that are interacting with it are low. As we discussed above, our system will eventually radiate most of its energy away and therefore the system is low energy enough that the bound state can be treated as a spin 0 particle. At this point the spin-spin force switches off because the two neutrinos in each bound state cancel each other. If neutrinos did not make spin 0 composite system they would obey the Pauli exclusion principle which will put a limit on the number of neutrinos that can exist in a given volume. This is why in the standard cosmology where no spin-spin force exist, neutrinos can not effectively participate in galaxy formations even if they are cold \cite{Tremaine}.

\section{Conclusions}
We have presented a cosmological model---SM-LGT---that makes lesser assumptions than the SM-GR-$\Lambda$-CDM and yet can explain the same observations. First, neutrinos as massive as 1eV can be the cold dark matter while in the SM-GR-$\Lambda$-CDM one needs to go beyond the standard model of particle physics to include the dark particles. Second, no dark energy is needed to explain the accelerating expansion of the universe in the late times and a decelerating universe spontaneously shifts to an accelerating one while in the SM-GR-$\Lambda$-CDM one needs to assume an extremely fine-tuned cosmological constant. Third, in our model the vacuum energy does not gravitate while SM-GR-$\Lambda$-CDM can not explain any of the current observations if one inserts the predicted value of the vacuum energy into the $\Lambda$. Fourth, LGT in the SM-LGT model has a much better UV behavior than GR in the SM-GR-$\Lambda$-CDM because the coupling constant in the latter has -2 dimensions in the units of mass but in the former it is dimensionless. 

In the presented cosmological model, the expansion of the universe is not very sensitive to the amount of radiation and therefore in order to explain observations of the primordial light elements and the spectrum of the CMB, neutrinos are not needed to be hot. On the other hand, besides the mass-generated gravitational force, there exists a long-range spin-generated force in our cosmological model that can keep neutrinos in thermal contact until they are cold.
Assuming that neutrinos are the cold dark matter, we place a lower bound of $10^{-7}$ on the dimensionless coupling constant of LGT. This is small enough to pass the upper bound on the constant that comes from the precision tests of the electroweak theory.


\end{document}